\documentclass[5p,sort&compress,numbers]{elsarticle}
\usepackage{epstopdf}
\usepackage{cuted}
\usepackage{subfigure,graphicx}
\usepackage{float}
\usepackage{amsmath, amsfonts, amssymb}
\usepackage{amsthm}
\usepackage{epsf}
\usepackage{amsfonts}
\usepackage{stmaryrd}
\usepackage{amssymb}
\usepackage{leftidx}
\usepackage{color}
\usepackage{mathtools}
\usepackage{placeins}
\usepackage{booktabs}
\usepackage{enumitem}
\usepackage{caption}
\usepackage{multirow}
\usepackage{stackengine}
\usepackage[utf8]{inputenc}
\usepackage[english]{babel}
\usepackage{color}
\usepackage{hyperref}

\theoremstyle{plain}

\def\bfu{{\bf u}}

\def\bfx{{\bf x}}

\def\bfE{{\bf E}}

\def\bfS{{\bf S}}

\def\bfX{{\bf X}}

\def\bfs{{\bf s}}

\def\e{\varepsilon}

\newcommand\sts{s_{\texttt{ts}}}

\newcommand\shs{s_{\texttt{hs}}}
\newcommand\sbs{s_{\texttt{bs}}}

\long\def\symbolfootnote[#1]#2{\begingroup%
\def\thefootnote{\fnsymbol{footnote}}\footnote[#1]{#2}\endgroup}

\begin{document}
\begin{frontmatter}

\title{Nucleation of fracture: The first-octant evidence against \\ classical variational phase-field models}

\author{Farhad Kamarei}
\ead{kamarei2@illinois.edu}

\author{John E. Dolbow}
\ead{jdolbow@duke.edu}

\author{Oscar Lopez-Pamies}
\ead{pamies@illinois.edu}

\address{Department of Civil and Environmental Engineering, University of Illinois, Urbana--Champaign, IL 61801, USA }

\address{Department of Mechanical Engineering, Duke University, Durham, NC 27708, USA \vspace{0.2cm}}

\vspace{1cm}

\begin{abstract}

\vspace{0.25cm}

As a companion work to \cite{LPDFL25}, this Note presents a series of simple formulae and explicit results that illustrate and highlight why classical variational phase-field models cannot possibly predict fracture nucleation in elastic brittle materials. The focus is on ``tension-dominated'' problems where all principal stresses are non-negative, that is, problems taking place entirely within the first octant in the space of principal stresses.

\vspace{0.3cm}

\keyword{Material strength; Energy methods; Configurational forces; Continuum balance principles}
\endkeyword

\end{abstract}

\end{frontmatter}

\section{Introduction}\label{Sec:IntroMain}

In spite of the evidence laid out against them in \cite{KFLP18,KLP20,KBFLP20}, classical variational phase-field models of fracture, or variational phase-field models for short,\footnote{By variational phase-field models we mean phase-field models of fracture that $\Gamma$-converge to the variational theory of brittle fracture \cite{Francfort98}. Many other phase-field models of fracture are variational but do not $\Gamma$-converge to that theory; see, e.g., \cite{Lorentz11,Iurlano16,Wu17,LDLP24}.} continue to be used and pursued in an attempt to describe fracture nucleation in elastic brittle materials. In this context, the work recently presented in \cite{LPDFL25} has provided a comprehensive review of the existing evidence that settles that such a class of models cannot possibly describe --- and hence predict --- fracture nucleation in general. As a companion work to \cite{LPDFL25}, this Note presents a series of simple formulae and explicit results aimed at illustrating and highlighting why this is the case. 

The focus of this Note is on ``tension-dominated'' problems where all principal stresses are non-negative, that is, problems taking place entirely within the first octant in the space of principal stresses. In this octant, virtually all variational phase-field models with energy splits that have been proposed in the literature reduce to the corresponding base models without energy split. This is so because energy splits have primarily been pursued to deal with the unphysical results produced by variational phase-field models in the presence of compressive strains and/or compressive stresses, and \emph{not} with the equally unphysical results that these models can produce when the stresses are all tensile.

\section{Variational phase-field models for elastic brittle materials}\label{Sec: The P-F model}

We begin by introducing notation and recalling the variational phase-field models of fracture for elastic brittle materials. Throughout, attention is restricted to isotropic linear elastic brittle materials and quasi-static loading conditions.

Consider a body made of an isotropic linearly elastic brittle material with elastic energy density $W(\bfE)$, strength surface $\mathcal{F}(\bfS)=0$, and toughness, or critical energy release rate, $G_c$ that, initially, at time $t=0$, occupies the open bounded domain $\Omega_0\subset \mathbb{R}^3$. We denote the boundary of the body by $\partial\Omega_0$ and identify material points by their initial position vector $\bfX\in \Omega_0$.

The body is subjected to a displacement $\overline{\bfu}(\bfX,t)$ on a part $\partial \Omega_0^{\mathcal{D}}$ of the boundary, and a surface force (per unit undeformed area) $\overline{\bfs}(\bfX,t)$ on the complementary part $\partial \Omega_0^{\mathcal{N}}=\partial \Omega_0\setminus\partial \Omega_0^{\mathcal{D}}$. In response to these stimuli --- both of which are assumed to be applied monotonically and quasi-statically in time --- the position vector $\bfX$ of a material point in the body will move to a new position specified by $\bfx=\bfX+\bfu(\bfX,t)$, where $\bfu(\bfX,t)$ is the displacement field. We write the associated strain at $\bfX$ and $t$ as $\bfE(\bfu)=\frac{1}{2}\left(\nabla\bfu+\nabla\bfu^T\right)$.

In addition to the deformation, the applied boundary conditions may result in the nucleation and subsequent propagation of cracks in the body. We describe such cracks in a regularized fashion via the phase field $v=v(\bfX,t)$ taking values in the range $[0, 1]$. 

According to the variational phase-field models (see, e.g., \cite{Bourdin00,Tanne18,Maurini24}), making use of the notation $\overline{\bfu}_k(\bfX)=\overline{\bfu}(\bfX,t_k)$ and $\overline{\bfs}_k(\bfX)=\overline{\bfs}(\bfX,t_k)$, the displacement field $\bfu_k(\bfX)=\bfu(\bfX,t_k)$ and phase field $v_k(\bfX)=v(\bfX,t_k)$ at any material point $\bfX \in \overline{\Omega}_0=\Omega_0\cup\partial \Omega_0$ and at any given discrete time $t_k\in\{0=t_0,t_1,...,t_m,t_{m+1},...,t_M=T\}$ are determined by a Nash minimizing pair  $(\bfu^\e_k, v^\e_k)$, subject to $\bfu^\e_k=\overline{\bfu}_k$ on $\partial \Omega_0^{\mathcal{D}}$ and $0\leq v^\e_k\leq v^\e_{k-1}\leq 1$, of the energy functional
\begin{align}\label{W-Split}
&\mathcal{E}^\varepsilon(\bfu_k,v_k):=\displaystyle\int_{\Omega_0} \left(g(v_k) W^+(\bfE(\bfu_k))+W^-(\bfE(\bfu_k))\right){\rm d}\bfX-\nonumber\\
&\displaystyle\int_{\partial \Omega_0^{\mathcal{N}}}\overline{\textbf{s}}_k\cdot\bfu_k\,{\rm d}\bfX +
\dfrac{3\,G_c}{8} \int_{\Omega_0}\left(\dfrac{1-v_k}{\varepsilon}+\varepsilon\nabla v_k\cdot\nabla v_k\right)\,{\rm d}\bfX,
\end{align}
where $\e>0$ is a regularization length, $g(v)$ is a function such that $g(0)=0$, $g(1)=1$, and where $W^+(\bfE)$ and $W^-(\bfE)\geq 0$ stand for any ``tensile'' and ``compressive'' parts of choice from the split $W(\bfE)=W^+(\bfE)+W^-(\bfE)$ of the elastic energy density. Specifically, by a Nash minimizing pair  $(\bfu^\e_k, v^\e_k)$ we mean the minimization pair $(\bfu^\e_k, v^\e_k)$ that is generated \emph{not} by global minimization but by alternating minimization. This choice of minimization is consistent with the fact that, while not convex, the energy functional (\ref{W-Split}) is separably convex in its arguments \cite{LPDFL25}.

\section{The strength surface generated by the variational phase-field models in the first octant}\label{Sec: The evidence}

As reviewed in \cite{LPDFL25}, and as first shown in \cite{KBFLP20}, when the body is subjected to a state of spatially uniform stress $\bfS$, the variational phase-field models (\ref{W-Split}) may predict that fracture nucleates at some critical value of $\bfS$. If and when fracture nucleation is predicted depends on the value of the regularization length $\e$, the specific type of energy split $W(\bfE)=W^+(\bfE)+W^-(\bfE)$, as well as on the choice of the degradation function $g(v)$. 

For the prominent case when $g(v)=v^2$, the stress $\bfS={\rm diag}(s_1\geq 0,s_2\geq 0,s_3\geq 0)$ is in the first octant in the space of principal stresses $(s_1,s_2,s_3)$, and $W^-(\bfE)=0$ so that $W(\bfE)=W^+(\bfE)$, Kumar et al. \cite{KBFLP20} showed that the resulting variational phase-field model does indeed predict fracture nucleation at some critical value of $\bfS$. The set of all such critical stresses $\bfS$ defines the following surface in stress space:
\begin{equation}\label{F-AT1}
\mathcal{F}^{{\texttt{AT}}_{1}}(\bfS)=\dfrac{\mathcal{J}_2}{\mu}+\dfrac{\mathcal{I}_1^2}{9\kappa}-\dfrac{3 G_c}{8\varepsilon}=0,\;\; s_1,s_2,s_3\geq 0.
\end{equation}
Here, $\mathcal{I}_1=s_1+s_2+s_3$, $\mathcal{J}_2=\frac{1}{3}(s_1^2+s_2^2+s_3^2-s_1 s_2-s_1s_2-s_2 s_3)$, and $\mu$ and $\kappa$ denote the shear and bulk moduli of the material. For later convenience, we recall that $\mu$ and $\kappa$ are given in terms of the Young's modulus $E$ and the Poisson's $\nu$ ratio by $\mu=E/(2(1+\nu))$ and $\kappa=E/(3(1-2\nu))$.

Virtually all energy splits that have been proposed in the literature (see, e.g., the reviews included in \cite{Wick22,Perego24}) are such that $W^-(\bfE)=0$ when $\bfS={\rm diag}(s_1\geq 0,s_2\geq 0,s_3\geq 0)$. As a result, the surface (\ref{F-AT1}) can be viewed as the strength surface that is generated in the first octant of principal stresses by any variational phase-field model (\ref{W-Split}) with $g(v)=v^2$.

Note, in particular, that for uniaxial tension, when $\bfS={\rm diag}(s>0,0,0)$, the surface (\ref{F-AT1}) predicts the uniaxial tensile strength 
\begin{equation}\label{sts-AT1}
\sts^{{\texttt{AT}}_{1}}=\sqrt{\dfrac{3 G_c E}{8\varepsilon}},
\end{equation}
while for equi-biaxial and hydrostatic tension, when $\bfS={\rm diag}(s>0,s>0,0)$ and $\bfS={\rm diag}(s>0,s>0,s>0)$, it predicts the equi-biaxial and hydrostatic tensile strengths
\begin{equation}\label{sbs-shs-AT1}
\sbs^{{\texttt{AT}}_{1}}=\sqrt{\dfrac{3 G_c E}{16(1-\nu)\varepsilon}}\quad {\rm and}\quad \shs^{{\texttt{AT}}_{1}}=\sqrt{\dfrac{G_c E}{8(1-2\nu)\varepsilon}}.
\end{equation}

Clearly, the surface (\ref{F-AT1}) is not an independent material property that can be chosen to match the actual strength surface $\mathcal{F}(\bfS)=0$ of the material; recall that $\mathcal{F}(\bfS)=0$ is potentially any star-shaped surface in stress space containing 0 in its interior \cite{KLP20,KBFLP20,LPDFL25}. In particular, the surface (\ref{F-AT1}) is subordinate to the elasticity and toughness of the material, even if $\varepsilon$ is viewed as a material length scale. This is in contradiction with experimental observations and the fundamental reason why variational phase-field models cannot possibly describe fracture nucleation \cite{LPDFL25}.

When considered as a material length scale, $\varepsilon$ is the sole tunable parameter in (\ref{F-AT1}), one whose value can be selected so that the strength surface (\ref{F-AT1}) is forced to match the actual strength surface $\mathcal{F}(\bfS)=0$ of the material at a single point of choice in stress space. Making use of the popular prescription 
\begin{align}\label{eps-AT1}
\varepsilon=\dfrac{3 G_c E}{8\sts^2}
\end{align}
proposed in \cite{Tanne18}, where $\sts$ stands for the actual uniaxial tensile strength of the material, the strength surface (\ref{F-AT1}) specializes to
\begin{equation}\label{F-AT1-II}
\mathcal{F}^{{\texttt{AT}}_{1}}(\bfS)=2(1+\nu)\mathcal{J}_2+\frac{1-2\nu}{3}\mathcal{I}_1^2-\sts^2=0,\;\; s_1,s_2,s_3\geq 0,
\end{equation}
while relations (\ref{sts-AT1}) and (\ref{sbs-shs-AT1}) specialize to
\begin{equation}\label{sbs-shs-AT1-eps}
\sts^{{\texttt{AT}}_{1}}=\sts,\;\; \sbs^{{\texttt{AT}}_{1}}=\dfrac{\sts}{\sqrt{2(1-\nu)}},\;\;\shs^{{\texttt{AT}}_{1}}=\dfrac{\sts}{\sqrt{3(1-2\nu)}}.
\end{equation}
That is, the prescription (\ref{eps-AT1}) forces the uniaxial tensile strength $\sts^{{\texttt{AT}}_{1}}$ predicted by the variational phase-field model to agree identically with the actual uniaxial tensile strength $\sts$ of the material. At the same time, it subordinates all other remaining points on the surface (\ref{F-AT1-II}) to $\sts$ and the Poisson's ratio $\nu$ of the material. Such a subordination is unphysical.

\begin{figure}[t!]
   \centering \includegraphics[width=2.7in]{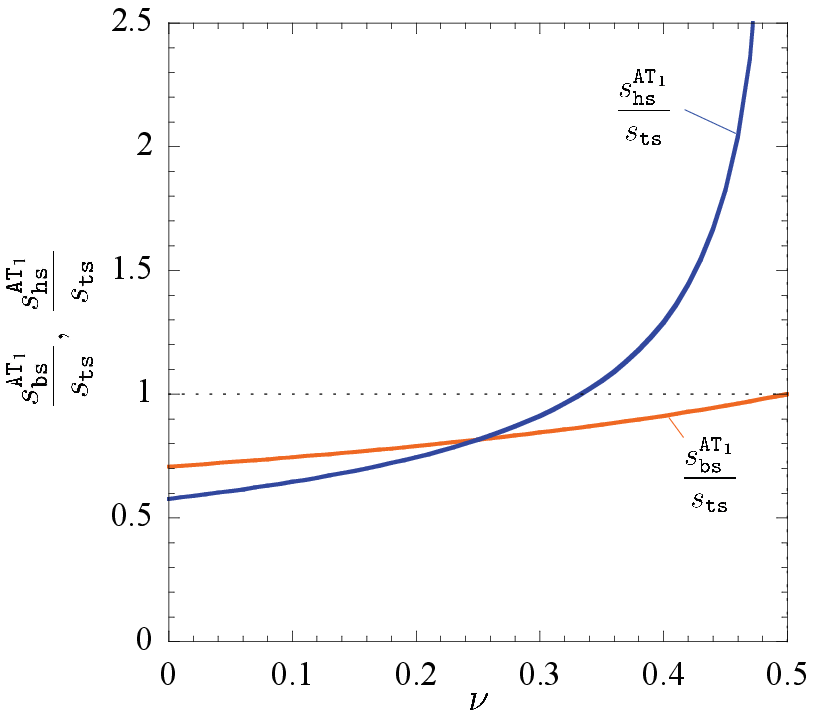}
   \vspace{0.2cm}
   \caption{Plots illustrating the unphysical dependence of the predicted equi-biaxial tensile strength (\ref{sbs-shs-AT1-eps})$_2$ and the hydrostatic tensile strength (\ref{sbs-shs-AT1-eps})$_3$ on the Poisson's ratio $\nu$ of the material. The results are plotted normalized by the actual uniaxial tensile strength $\sts$ of the material. }\label{Fig1}
\end{figure}

It proves instructive to visualize how the strength surface (\ref{F-AT1-II}) depends on $\nu$. To this end, Fig. \ref{Fig1} plots the equi-biaxial tensile strength (\ref{sbs-shs-AT1-eps})$_2$ and the hydrostatic tensile strength (\ref{sbs-shs-AT1-eps})$_3$, normalized by $\sts$, as a function of $\nu$. Both $\sbs^{{\texttt{AT}}_{1}}$ and $\shs^{{\texttt{AT}}_{1}}$ are seen to increase monotonically with increasing values of $\nu$, even though, again, the actual strength of a material is independent of its elasticity. It is also interesting to note that $\sbs^{{\texttt{AT}}_{1}}\leq \sts$ for all $\nu$, while $\shs^{{\texttt{AT}}_{1}}\leq \sbs^{{\texttt{AT}}_{1}}$ for $\nu\leq1/4$ and $\shs^{{\texttt{AT}}_{1}}\leq \sts$ for $\nu\leq1/3$. For $\nu>1/3$, not only $\shs^{{\texttt{AT}}_{1}}> \sts$ but $\shs^{{\texttt{AT}}_{1}}= +\infty$ at $\nu=1/2$. This behavior is nonsensical. 

%
%%%%%%%%%%%%%%%%%%%%%%%%%%%%%%%%%%%%%%%%%%%%%%%%%%%%%%%%%%%%%%%%%%%%%%%%%%%%%%
%%%%%%%%%%%%%%%%%%%%%%%%%%%%%%%%%%%%%%%%%%%%%%%%%%%%%%%%%%%%%%%%%%%%%%%%%%%%%%
%

\section{Final comments}\label{Sec: comments}

Per the large body of experimental observations that have been gathered for over a century, as reviewed in \cite{LPDFL25}, there are three necessary requirements that any phase-field model, be it variational or not, must satisfy if it is to potentially describe fracture nucleation in elastic brittle materials. These are:
\begin{enumerate}[label=\emph{\roman*}.]

\item{Accounting for the elastic energy density $W(\bfE)$, the strength surface $\mathcal{F}(\bfS)=0$, and the toughness $G_c$ of the material, whatever these properties may be;}

\item{Localization of the phase field $v$ whenever a macroscopic piece of the material is subjected to any uniform stress $\bfS$ that exceeds the strength surface $\mathcal{F}(\bfS)=0$ of the material; and}

\item{Having the Griffith energy competition as a descriptor of fracture nucleation from the front of a large pre-existing crack.}

\end{enumerate}
Failure to satisfy any of these requirements would prevent the model from describing fracture nucleation even in the simplest of scenarios, that is, under a spatially uniform stress and/or from large pre-existing cracks. 

As illustrated by results presented in the preceding section, variational phase-field models fail to satisfy the requirement $i$, since they cannot account for the strength surface $\mathcal{F}(\bfS)=0$ as an independent material property.

A class of phase-field models that satisfies all the above three requirements is that introduced in \cite*{KFLP18,KLP20,KBFLP20}; see, also, \cite{KKLP24,LDLP24}. The distinguishing feature of this class of models is that they account for the strength surface $\mathcal{F}(\bfS)=0$ via a driving force $c_\texttt{e}$ in the evolution equation for the phase field $v$. This driving force is designed in a manner such that the requirements $ii$ and $iii$ are satisfied.

%
%%%%%%%%%%%%%%%%%%%%%%%%%%%%%%%%%%%%%%%%%%%%%%%%%%%%%%%%%%%%%%%%%%%%%%%%%%%%%%
\begin{figure}[b!]
   \centering \includegraphics[width=2.62in]{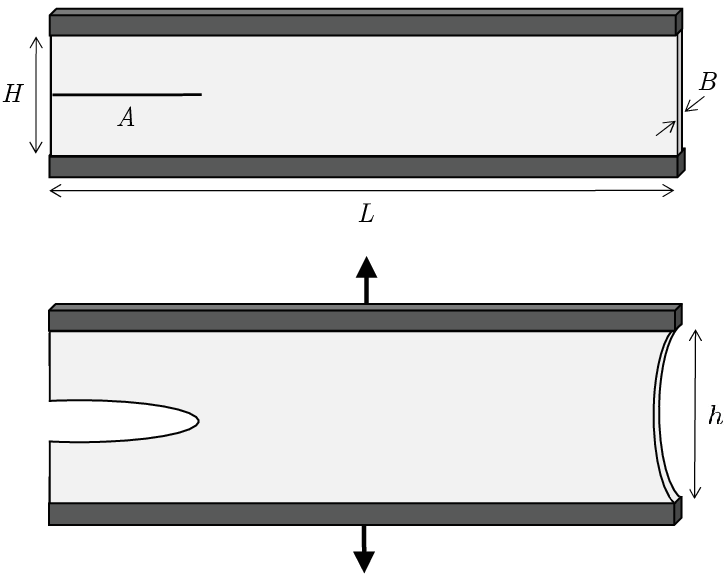}
   \vspace{0.2cm}
   \caption{Schematics of the ``pure-shear'' fracture test in the initial configuration and in a deformed configuration at an applied deformation $h$.}\label{Fig2}
\end{figure}
%%%%%%%%%%%%%%%%%%%%%%%%%%%%%%%%%%%%%%%%%%%%%%%%%%%%%%%%%%%%%%%%%%%%%%%%%%%%%%
%

An important lesson that emerged from \cite*{KFLP18,KLP20,KBFLP20} is that any willy-nilly attempt to account for the strength surface $\mathcal{F}(\bfS)=0$ will impact how the resulting model predicts fracture nucleation from the front of a large pre-existing crack. Put differently, even if one manages to correctly account for the strength surface $\mathcal{F}(\bfS)=0$, additional steps must be taken to ensure that the model remains consistent with the Griffith energy competition as a descriptor of fracture nucleation from the front of a large pre-existing crack. This is because fracture nucleation predictions by a phase-field model, with a finite value for regularization length $\varepsilon$, are generally strongly dependent on \emph{all} of the specifics of the model, including the choice of degradation function $g(v)$. Below, we illustrate this key point by making use of the two different degradation functions \cite{Landis23,Larsen23}
\begin{equation}\label{Chad-g}
g(v)=\alpha\left(1-\left(\dfrac{\alpha-1}{\alpha}\right)^{v^2}\right)
\end{equation}
and
\begin{equation}\label{Chris-g}
g(v)=\left\{\begin{array}{ll}1 & {\rm if}\; v\geq\beta \vspace{0.2cm}\\
v^2 & {\rm if}\; v<\beta\end{array}\right. ,
\end{equation}
where $\alpha> 1$ and $0<\beta\leq 1$, in the variational phase-field model (\ref{W-Split}) to predict fracture nucleation from a large pre-existing crack in a ``pure-shear'' fracture test. Consistent with the focus of this Note, the stresses around the crack front (where fracture nucleation or crack growth occurs) in such a test are all within the first octant in the space of principal stresses. We remark that both degradation functions (\ref{Chad-g}) and (\ref{Chris-g}) are such that the resulting variational phase field model (\ref{W-Split}), with $W^{-}(\bfE)=0$,  $\Gamma$-converges to the variational theory of brittle fracture \cite{Francfort98}. While (\ref{Chad-g}) was introduced to be able to deal with large structures \cite{Landis23}, the prescription (\ref{Chris-g}) was suggested as a simple modification to prevent undesirable fracture nucleation \cite{Larsen23}. For our purposes here, we use them simply as representative examples of degradation functions that generalize the basic choice $g(v)=v^2$. Indeed, observe that (\ref{Chad-g}) reduces to $g(v)=v^2$ when $\alpha=+\infty$, while (\ref{Chris-g}) reduces to $g(v)=v^2$ when $\beta=1$.

Because of its experimental convenience together with the fact that its analysis can be carried out explicitly, the ``pure-shear'' fracture test is one of a handful of tests preferred by practitioners to measure the toughness $G_c$ of materials. As schematically depicted by Fig. \ref{Fig2}, the test makes use of a specimen in the form of a plate of initial height $H$, much smaller thickness $B\ll H$, and much larger length $L\gg H$ that contains a large pre-existing crack, of initial size $A>H$, on one of its sides along its centerline. The specimen is clamped on its top and bottom and subjected to a prescribed deformation $h$ between the grips. For an isotropic linear elastic brittle material, the critical value $h_{cr}$ of the applied deformation $h$ at which the pre-existing crack will start growing according to the Griffith energy competition can be accurately estimated from the formula \cite{RT53}
\begin{equation}\label{ecr}
h_{cr}=(1+e_{cr})H\quad {\rm with}\quad e_{cr}=\sqrt{\dfrac{2(1-\nu^2)G_c}{H E}}
\end{equation}
in terms of the critical value $e_{cr}$ of the global strain $e=(h-H)/H$.

Tables \ref{Table1} and \ref{Table2} report the error in the critical global strains predicted by the variational phase-field models (\ref{W-Split}) with $W^{-}(\bfE)=0$ and degradation functions (\ref{Chad-g}) and (\ref{Chris-g}) at which fracture nucleates in  ``pure-shear'' fracture tests for various values of the parameters $\alpha$ and $\beta$. In particular, the results in Table \ref{Table1} correspond to four of the values examined in \cite{Landis23}, namely, $\alpha=1.0148,1.017,1.1,200$. The results in Table \ref{Table2} correspond to $\beta=0.4,0.65,0.9,1$, the last of which, again, amounts to nothing more than the basic choice $g(v)=v^2$ for the degradation function. All the results pertain to simulations for specimens of initial height $H=5$ mm, length $L=50$ mm, thickness $B=1$ mm, and crack length $A=10$ mm that are made of a material with elastic constants and toughness that are representative of titania, to wit, $E=250$ GPa,  $\nu=0.29$, and $G_c=36$ N/m; see Section 4.3 in \cite{LPDFL25}. For such specimens and material constants, expressions (\ref{ecr}) yield $e_{cr}=0.2296\times 10^{-3}$ and $h_{cr}=5.0012$ mm. All the simulations are carried out with the same regularization length, $\varepsilon=0.33$ mm, and the same finite-element mesh of size $\texttt{h}=\varepsilon/5=0.066$ mm.

The results in Tables  \ref{Table1} and \ref{Table2} call for the following observations. The prediction by the variational phase-field model (\ref{W-Split}) with the basic choice $g(v)=v^2$ for the degradation function is practically in perfect agreement with the sharp solution. This is because such a model features a fast convergence as $\varepsilon\searrow 0$ and, for the problem at hand, the value $\varepsilon=0.33$ mm is already sufficiently small to yield agreement with the sharp solution (\ref{ecr}). By contrast, the variational phase-field models (\ref{W-Split}) with degradation functions (\ref{Chad-g}) and (\ref{Chris-g}) exhibit slower convergence as $\varepsilon\searrow 0$. In particular, $\varepsilon=0.33$ mm is not sufficiently small for these models to yield agreement with the sharp solution. The disagreement increases significantly with decreasing values of $\alpha$ and $\beta$. These results illustrate one of the difficulties that one must face in constructing computationally tractable phase-field models of fracture that are aimed at describing fracture nucleation in general.

\begin{table}[H]\centering
\caption{Error in the critical global strains ${}^{\alpha}e_{cr}$ predicted by the variational phase-field model (\ref{W-Split}) with degradation function (\ref{Chad-g}) for various values of the parameter $\alpha$. All results pertain to the same regularization length ($\varepsilon=0.33$ mm).}
\begin{tabular}{r|cccc}
\toprule
$\alpha$  & $1.0148$ & $1.017$ & $1.1$ & $200$ \\
\hline
$({}^{\alpha}e_{cr}-e_{cr})/e_{cr}$  & $34\%$ & $33\%$ & $31\%$ & $13\%$\\
\bottomrule
\end{tabular} \label{Table1}
\end{table}
\begin{table}[H]\centering
\caption{Error in the critical global strains ${}^{\beta}e_{cr}$ predicted by the variational phase-field model (\ref{W-Split}) with degradation function (\ref{Chris-g}) for various values of the parameter $\beta$. All results pertain to the same regularization length ($\varepsilon=0.33$ mm).}
\begin{tabular}{r|cccc}
\toprule
$\beta$  & $0.4$ & $0.65$ & $0.9$ & $1$\\
\hline
$({}^{\beta}e_{cr}-e_{cr})/e_{cr}$  & $222\%$ & $39\%$ & $22\%$ & $4\%$\\
\bottomrule
\end{tabular} \label{Table2}
\end{table}

\section*{Acknowledgements}

This work was supported by the National Science Foundation through the Grants CMMI--2132528, CMMI--2132551. This support is gratefully acknowledged.

\bibliographystyle{unsrtnat}
\bibliography{References}

\end{document}